\documentclass[epj]{svjour}
\usepackage{graphics}
\usepackage{braket}
\usepackage{bm}
\usepackage{cite}
\usepackage{amsfonts} 
\usepackage{amsmath}
\usepackage{hyperref}
\usepackage{breqn}
\begin{document}
\title{First-principles dynamics of electrons and phonons}
\author{Marco Bernardi\inst{1}
}                     
%
%
\institute{Department of Applied Physics and Materials Science, Steele Laboratory, California Institute of Technology, Pasadena, CA 91125, USA.}
%
%
\abstract{First-principles calculations combining density functional theory and many-body perturbation theory can provide microscopic insight into the dynamics of electrons and phonons in materials.
We review this theoretical and computational framework, focusing on perturbative treatments of scattering, dynamics and transport of coupled electrons and phonons.
We discuss application of these first-principles calculations to electronics, lighting, spectroscopy and renewable energy.
%
} 
\maketitle
\section{Introduction}
\label{intro}
Solid-state technologies depend crucially on the dynamics of electrons, phonons and excited states. For example, charge transport in electronic and optoelectronic devices is limited by the scattering of charge carriers with phonons and defects \cite{Ziman,Mahan}, and phonon-phonon scattering controls heat transport and thermoelectric processes \cite{broido-01,gchen}. The efficiency of light-emitting devices depends on the fraction of excited electrons recombining radiatively \cite{LED} as opposed to non-radiatively \cite{nonrad-01,nonrad-02}, e.g., by multi-phonon emission, Auger processes, or defect trapping. Finally, spintronic devices \cite{spintronics} and solid-state qubits proposed for quantum computing \cite{qubit-01} rely on long-lived spin populations achieved by suppressing decoherence effects \cite{spin-01,qubit-02}. These dynamical processes take place on a femtosecond to nanosecond timescale, and as such are challenging to study experimentally \cite{Shah}. Computational approaches can provide new insight into this ultrafast dynamics in materials.\\
\indent
For the past few decades, first-principles calculations have focused on computing the \textit{energetics} of electrons, phonons and excited states. For example, computations of band gaps \cite{Hybertsen,Rubio}, optical spectra \cite{Rohlfing}, and phonon dispersions \cite{Baroni} have been a central focus of the ab initio community. However, the \textit{dynamics} of electrons and phonons is at least as important as the energetics to understand materials and devices. 
Analytical theories of the interactions among electrons, phonons, defects and excited states have been investigated extensively; comprehensive reviews exist on the subject \cite{Ziman,Mahan,Ridley}. 
%
%
Application of this theoretical framework has been limited by the lack of accurate approaches to compute such interactions quantitatively. Recent advances in first-principles calculations 
make it possible to compute electron and phonon interactions without employing heuristic parameters, thus enabling new discoveries in ultrafast dynamics and transport.\\
\indent
First-principles calculations of electron and phonon dynamics hold a unique promise for broad scientific impact.
Research benefitting from these novel approaches include device testing and manufacturing, both in industry and academia, as well as electronics, optoelectronics, and renewable energy (e.g., solar cells and thermoelectrics). 
In addition, calculations of the dynamics of excited electrons are crucial to advance ultrafast spectroscopy. The future success of tabletop and large-scale spectroscopy facilities, such as the free electron laser for X-ray time-resolved spectroscopy \cite{FEL,LCLS}, critically depend on computational tools able to microscopically interpret sophisticated experiments probing matter at increasingly short timescales.\\
\indent
This article discusses first-principles perturbative computations of electron and phonon dynamics, with the aim of bridging the gap between textbook treatments and the current literature as well as collecting into one article several computational approaches. 
Recent trends in the application of this framework to materials and devices are reviewed.
%
%
\section{Approaches to first-principles dynamics}
First-principles calculations aimed at extending density functional theory (DFT) \cite{Martin} and related excited-state methods \cite{Rubio} to study electron and phonon dynamics are in a relatively early stage of development. 
In particular, two main families of approaches are being explored. 
The first is real-time time-dependent DFT \cite{Octopus}, which employs the Kohn-Sham Hamiltonian to self-consistently propagate the electronic wavefunction and charge density \cite{Castro}. This approach has been applied extensively to study electron dynamics in materials and interfaces \cite{Rubio-1,Rozzi,Prezhdo}, as reviewed in a recent article \cite{Isborn}.
%
%
The second approach, which is the focus of this article, employs many-body perturbation theory \cite{Mahan-main,Coleman} to compute the electron-phonon \mbox{(e-ph)}, electron-electron (e-e), phonon-electron (ph-e), and phonon-phonon (ph-ph) interactions and scattering processes from first principles. Combined with the Boltzmann transport equation \cite{Mahan} or the Kadanoff-Baym equations \cite{Kadanoff,Jauho,Stefanucci}, this approach enables studies of transport and dynamics in materials. Additional interactions involving defects, spin, excitons, and various excited states are also being actively investigated, but will not be discussed here.\\
\indent
%
%
The Fermi golden rule (FGR) is the key tool to compute the rates and timescales of electron and phonon scattering processes. The FGR provides an intuitive understanding of scattering in terms of the matrix elements of the perturbation potential and the phase space of available final states: 
\begin{equation}
\Gamma_{\alpha_i} = \tau^{-1}_{\alpha_i} = \frac{2\pi}{\hbar} \sum_{\alpha_f} |M_{\alpha_i,\alpha_f}|^2 \delta\left({E_{\alpha_i} - E_{\alpha_f}}\right)
\label{fgr}
\end{equation}
where $\Gamma_{\alpha_i}$ is the scattering rate and $\tau_{\alpha_i}$ its inverse, the relaxation time (RT), for an initial state $i$ with quantum numbers $\alpha_i$ to scatter into final states $f$ with quantum numbers $\alpha_f$. 
The energy $E$ and momentum are conserved in the scattering process, and the scattering is induced by the matrix element $M$ of the perturbation potential $H'$ coupling the initial and final states, $\psi_{\alpha_i}$ and $\psi_{\alpha_f}$ respectively: 
\begin{equation}
M_{\alpha_i,\alpha_f} = \braket{\psi_{\alpha_i} | H' | \psi_{\alpha_f}}
\label{me}
\end{equation}
Here, we focus on solid state calculations, and assume that the Born-Oppenheimer approximation is valid, so that the matrix elements of interest are those associated with scattering processes among electrons and phonons.
In the language of many-body theory, the FGR approach in eq. \ref{fgr} corresponds to computing the imaginary part of the self-energy, $\mathrm{Im}\Sigma$, within the lowest-order of perturbation theory in the given interaction. 
For electrons and phonons, respectively, we obtain for the scattering rates and RTs: 
\begin{align}
\label{ex}
& \Gamma^{\mathrm{e-x}}_{n\bf{k}} \! =\! \left( \tau^{-1}_{n\bf{k}} \right)^{\mathrm{e-x}}= \frac{2}{\hbar} \mathrm{Im}\Sigma_{n\bf{k}}^{\mathrm{e-x}} \\
\label{phx}
& \Gamma^{\mathrm{ph-x}}_{\nu\bf{q}} \! =\! \left( \tau^{-1}_{\nu \mathbf{q}} \right)^{\mathrm{ph-x}}= \frac{2}{\hbar} \mathrm{Im}\Sigma_{\nu \mathbf{q}}^{\mathrm{ph-x}}
\end{align}
For each of the electron (e-ph and e-e) and phonon (ph-e and ph-ph) interactions, denoted as e-x in eq. \ref{ex} and ph-x in eq. \ref{phx} respectively, the corresponding self-energy diagrams are shown in Fig. 1, together with the scattering processes visualized with the aid of the electronic bandstructure and phonon dispersions. Here and throughout the article, $n$ is the band index and $\mathbf{k}$ the Brillouin zone (BZ) crystal momentum of electron Bloch states, while $\nu$ and $\mathbf{q}$ are the phonon branch index and wavevector, respectively. For first-principles calculations of electron and phonon scattering rates, the sum over final states in the FGR requires integration of the scattering matrix elements, bandstructure and phonon dispersions on fine BZ grids, making these calculations computationally challenging, as discussed below.\\
\indent
%
%
The rest of this paper is organized as follows: 
Section \ref{sec:1} discusses the interactions among electrons and phonons within many-body perturbation theory. Electron and phonon scattering rates and RTs are derived within the framework of ab initio calculations. 
Section \ref{sec:2} examines applications to dynamics and transport of carriers and phonons.\\
\indent
%
%
%
%
%
%
\begin{figure*}
\centering
\resizebox{0.9\textwidth}{!}{
\includegraphics{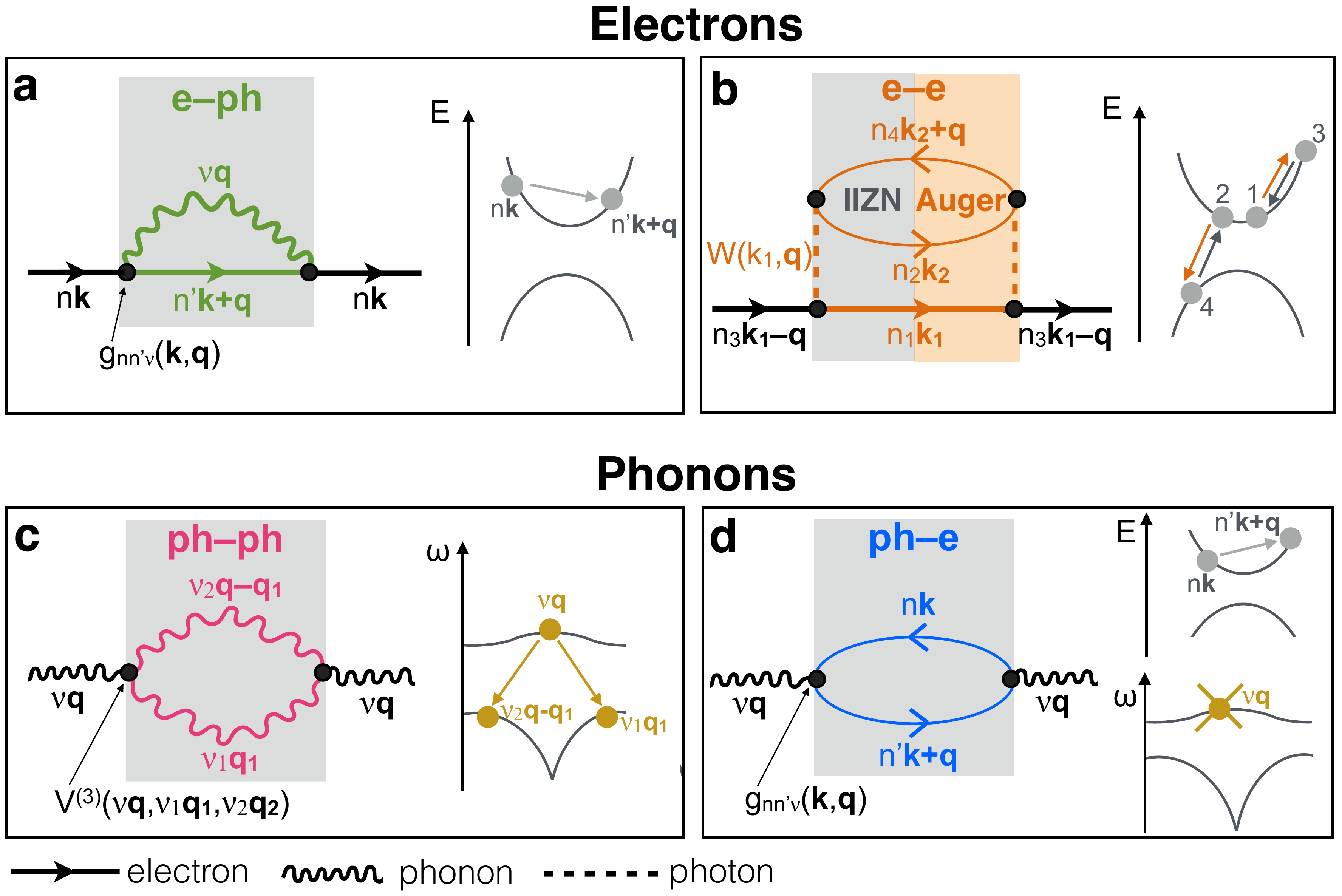}}
\caption{\textbf{Electron (upper panel) and phonon (lower panel) scattering processes.} Feynman diagrams for the e-ph (a), e-e (b), ph-ph (c) and ph-e (d) scattering processes. For all processes, the self-energy is shown in a shaded box, and the scattering processes for electrons (light gray) or phonons (dark yellow) are mapped onto the electronic bandstructure or phonon dispersions. For the ph-e interaction, the disappearance of a phonon is indicated with a cross. Note that in all cases, the self-energy is built by joining two half-diagrams representing microscopic scattering processes. For example, the e-ph self-energy diagram in (a) is obtained by joining the two diagrams for phonon emission (left half) and absorption (right half), and the e-e self-energy in (b) by joining the diagrams for the IIZN and $eeh$ Auger processes, which are labeled separately in gray and orange, respectively.}
\label{fig1} 
\end{figure*}
\section{Electron and phonon interactions}
\label{sec:1}
\subsection{Electron-phonon}
%
%
The e-ph interaction is a crucial ingredient to understanding carrier dynamics. For example, the energy loss rate of excited carriers \cite{Bernardi-Si,Bernardi-GaAs,Bernardi-Au} 
and the room temperature electrical conductivity in crystals with low impurity concentrations \cite{Mahan,Restrepo,Bernardi-transport} are both controlled by e-ph scattering. 
In DFT \cite{Martin}, the Kohn-Sham (KS) potential $V^{\mathrm{KS}}$, computed at the equilibrium atomic positions, is employed to obtain the electronic KS bandstructure $\epsilon_{n\mathbf{k}}$ and wavefunctions $\phi_{n\mathbf{k}}$. 
The vibrational motion displaces the nuclei (or ions, i.e., nuclei plus core electrons) from their equilibrium positions, thus perturbing the electronic states. This perturbation results in a temperature-dependent shift of the bandstructure and a finite electron lifetime, defined here as the e-ph RT.\\
\indent 
%
%
To derive the e-ph RT, the KS potential is first expanded in powers of the ion displacements $\{ \mathbf{u}_{is}\}$ away from the equilibrium position, where $s$ labels the atom and $i$ the unit cell in the Born-von Karman (BvK) supercell:
\begin{equation}
V^{\mathrm{KS}} (\{ \mathbf{u}_{is} \}) = V^{\mathrm{KS}}_0 + \sum_{is\alpha} \frac{\partial V^{\mathrm{KS}} }{ \partial \mathbf{u}_{is \alpha} }  \mathbf{u}_{is \alpha} + \mathcal{O}( \{\mathbf{u}_{is \alpha} \}^2)
\label{expan}
\end{equation}
where both $V^{\mathrm{KS}}_0$ and the derivatives of $V^{\mathrm{KS}}$ are computed at the equilibrium atomic positions.\footnote{The term of order $\mathcal{O}( \{\mathbf{u}_{is \alpha} \}^2)$ leads to the so-called Debye-Waller correction, but is not considered here since its lowest-order perturbation does not affect the RTs.} Here and throughout the paper, $\alpha = (x,y,z)$ are Cartesian coordinates.
%
Second quantization of the vibrational degrees of freedom is then carried out by expressing the ion displacements in terms of phonon annihilation and creation operators, $\hat{b}_{\nu \mathbf{q}}$ and $\hat{b}^{\dagger}_{\nu \mathbf{q}}$, respectively: 
\begin{equation}
\mathbf{\hat{u}}_{is\alpha} =  \sum_{\nu \mathbf{q}} \left( \frac{\hbar}{ 2 M_s \omega_{\nu \mathbf{q}} \mathcal{N}}   \right)^{\!1/2} \! \mathbf{e}^{s\alpha}_{\nu \mathbf{q}} \, e^{i \mathbf{q} \cdot \mathbf{R}_i} ( \hat{b}_{\nu \mathbf{q}} + \hat{b}^{\dagger}_{\nu -\mathbf{q}} )
\label{displacement}
\end{equation} 
where $\mathbf{e}^{s\alpha}_{\nu \mathbf{q}}$ is the $\alpha$-th component of the phonon eigenvector for the atom $s$, $\omega_{\nu \mathbf{q}}$ are phonon frequencies, $M_s$ atomic masses, $\mathbf{R}_i$ lattice vectors and $\mathcal{N}$ the number of unit cells in the BvK supercell.
%
%
Inserting eq. \ref{displacement} into eq. \ref{expan}, we obtain the operator for the e-ph perturbation potential:
\begin{equation}
\hat{V}^{\mathrm{KS}}(\{ \mathbf{u}_{is} \}) - \hat{V}^{\mathrm{KS}}_0 = \sum_{\nu \mathbf{q}} \left( \frac{\hbar}{ 2 \omega_{\nu \mathbf{q}} } \right)^{\!1/2} \Delta_{\nu \mathbf{q} } V^{\mathrm{KS}} ( \hat{b}_{\nu \mathbf{q}} + \hat{b}^{\dagger}_{\nu -\mathbf{q}} )
\label{pert-eph-01}
\end{equation}
where we defined generalized phonon-mode-resolved e-ph perturbations:
\begin{equation}
\Delta_{\nu \mathbf{q}} V^{\mathrm{KS}} =  \sum_{s\alpha} \frac{ \mathbf{e}^{s\alpha}_{\nu \mathbf{q}} }{\sqrt{M_s}} \! \sum_i \frac{ e^{i\mathbf{q} \cdot \mathbf{R}_i }}  { \sqrt{\mathcal{N}} } \frac{\partial V^{\mathrm{KS}} }{ \partial \mathbf{u}_{is \alpha} }
\label{mode-resolved}
\end{equation}
%
%
The first-principles e-ph perturbation Hamiltonian, $\hat{H}^{\mathrm{e-ph}}$, is then obtained through second quantization of the electronic states using electron annihilation and creation operators, 
$\hat{c}_{n \mathbf{k}}$ and $\hat{c}^{\dagger}_{n \mathbf{k}}$ respectively:
\begin{equation}
\begin{split}
&\hat{H}^{\mathrm{e-ph}} = \sum_{\substack{ n \mathbf{k}  \\ n' \mathbf{k}' } } \braket{\phi_{n'\mathbf{k'}} | \hat{V}^{\mathrm{KS}}( \{ \mathbf{\hat{u}}_{is} \}) - \hat{V}^{\mathrm{KS}}_0 | \phi_{n \mathbf{k}}} \hat{c}^\dagger_{n' \mathbf{k'}} \hat{c}_{n \mathbf{k}} \\
&\!\!\!\!\!\!\!\!=  \!\!\left( \frac{\hbar}{ 2 \omega_{\nu \mathbf{q}} } \right)^{\!\!\!1/2}\!\!\! \sum_{ \substack{ n n' \mathbf{k}  \\ \nu \mathbf{q}}} \!\! \braket{ \phi_{n'\mathbf{k}+\mathbf{q}} | \Delta_{\nu \mathbf{q}} V^{\mathrm{KS}} | \phi_{n \mathbf{k}}}  \hat{c}_{n' \mathbf{k} + \mathbf{q}}^\dagger \hat{c}_{n \mathbf{k}} ( \hat{b}_{\nu \mathbf{q}} + \hat{b}^{\dagger}_{\nu -\mathbf{q}} ) 
\label{heph}
\end{split}
\end{equation}
where we applied crystal momentum conservation in the second line of eq. \ref{heph}. The e-ph matrix elements, $g_{nn'\nu} ( \mathbf{k}, \mathbf{q})$, are thus defined as:
\begin{equation}
g_{nn'\nu} ( \mathbf{k}, \mathbf{q}) = \left( \frac{\hbar}{ 2 \omega_{\nu \mathbf{q}} } \right)^{\!1/2} \braket{ \phi_{n'\mathbf{k}+\mathbf{q}} | \Delta_{\nu \mathbf{q}} V^{\mathrm{KS}} | \phi_{n \mathbf{k}}}  
\label{eph-me}
\end{equation}
and the e-ph Hamiltonian is then put in the standard form, which extends the typical textbook treatment \cite{Bruus} by including multiple electronic bands and phonon branches:
%
%
\begin{equation}
\hat{H}^{\mathrm{e-ph}} =\sum_{\substack{ n n' \mathbf{k}  \\ \nu \mathbf{q}}} g_{nn'\nu} ( \mathbf{k}, \mathbf{q})  \hat{c}^\dagger_{n' \mathbf{k} + \mathbf{q}} \hat{c}_{n \mathbf{k}} ( \hat{b}_{\nu \mathbf{q}} + \hat{b}^{\dagger}_{\nu -\mathbf{q}} ) 
\label{eph-sq}
\end{equation}
%
%
%
%
The e-ph matrix elements can be computed with density functional perturbation theory (DFPT) \cite{Baroni,Deinzer}, for example using the Quantum ESPRESSO \cite{QE} code.\\
\indent
Starting from the e-ph Hamiltonian, the e-ph RTs can be derived in several ways. It is instructive to analyze them in detail, after stating the result upfront.
%
%
Because the e-ph Hamiltonian creates or annihilates one phonon, the lowest order of perturbation theory giving a non-zero self-energy is the second order, namely order $ \mathcal{O}( \left| g_{n n' \nu} (\mathbf{k},\!\mathbf{q}) \right|^2 )$ in the e-ph matrix elements. 
The e-ph scattering rates obtained within second order perturbation theory read:
%
%
\begin{equation}
\begin{split}
\!\!\!\!\!\!\!\!\!\!\!\!\!\!\!\!\! &\Gamma^{\mathrm{e-ph}}_{n\bf{k}} (T) \!=\! \frac{2\pi} {\hbar} \frac{1}{\mathcal{N}}\!\! \sum_{n' \nu \bf{q}}\! \left| g_{n n' \nu} (\mathbf{k},\!\mathbf{q}) \right|^2 \!  \left[  \left( \! N_{\nu \bf{q}} \!+ \!1 - \! f_{n'\bf{k}+\bf{q}} \right)\! \right. \\
& \!\!\!\! \left. \times \delta ( { E_{n\bf{k}}  \! -\! E_{n' \bf{k} \! + \bf{q}} \!-\! \hbar \omega_{\nu\bf{q}} }) \!+ \! \left( N_{\nu \bf{q}} \! + \!f_{n'\bf{k}+\bf{q}}  \right) \! \delta ( {E_{n\bf{k}} \! - E_{n' \bf{k} \!+\bf{q}} \!+\! \hbar \omega_{\nu,\bf{q}}})  \right]  
\end{split}
\label{eph}
\end{equation}
where their inverse, the RTs $\tau^{\mathrm{e-ph}}_{n \mathbf{k}} = (\Gamma^{\mathrm{e-ph}}_{n\bf{k}})^{-1}$, can also be computed. In eq. \ref{eph}, the e-ph matrix elements describe an electron in Bloch state $\ket{n\mathbf{k}}$ (with quasiparticle energy $E_{n\mathbf{k}}$) that scatters into state $\ket{n'\mathbf{k}+\mathbf{q}}$ with quasiparticle energy $E_{n'\mathbf{k}+\mathbf{q}}$ due to a phonon with branch index $\nu$, wavevector $\mathbf{q}$ and frequency $\omega_{\nu \mathbf{q}}$. The first and second terms in square brackets correspond to phonon emission and absorption, respectively, and the temperature dependence of the scattering rate stems from the electron and phonon occupation factors, $f_{n\mathbf{k}}$ and $N_{\nu \mathbf{q}}$ respectively, while the e-ph matrix elements are typically computed at zero temperature using DFPT. 
%
%
Note that the scattering rates and RTs are resolved for different bands and $\mathbf{k}$-points in first-principles calculations, thus providing rich microscopic information. 
%
%
To converge the e-ph RTs, interpolation of the e-ph matrix elements is necessary since the sum in eq. \ref{eph} requires $10^4$$-$$10^6$ $\mathbf{q}$-points to converge \cite{Bernardi-transport,Bernardi-Si,Bernardi-GaAs,Bernardi-Au}, an unrealistic task for direct DFPT calculations due to computational cost. An interpolation approach using maximally localized Wannier functions \cite{W90-code} has been recently developed \cite{EPW-method}; alternative interpolation schemes using localized basis sets are being investigated by the author. \\
\indent
%
A rigorous and general approach to derive eq. \ref{eph} consists of treating the e-ph interaction perturbatively, using the Feynman-diagram technique. The equilibrium electron Green's function $G_{n \mathbf{k}}(\tau)$ in imaginary time $\tau$ is first expanded using finite-temperature perturbation theory \cite{Bruus}:
\begin{equation}
\begin{split}
\!\!\!\!\!\!\!G_{n \mathbf{k}}(\tau) &\!=\! - \!\sum_{m=0}^{\infty}\! \frac{(-1)^m}{ m!}\!\! \int_0^\beta \!\!\!\! d\tau_1 \!\cdots\!\! \int_0^\beta \!\!\! \!d\tau_m \\ 
& \times \left< T_\tau \hat{H}^{\mathrm{e-ph}} (\tau_1) \cdots \hat{H}^{\mathrm{e-ph}} (\tau_m) \hat{c}_{n \mathbf{k}} (\tau) \hat{c}^\dagger_{n \mathbf{k}} (0) \!\right>_{\!\!C}
\label{Matsubara}
\end{split}
\end{equation}
\noindent
where $\beta = (k_B T)^{-1}$ is the inverse temperature, the electron and phonon creation and annihilation operators in $\hat{H}^{\mathrm{e-ph}}$ are in the interaction picture, 
and the subscript $C$ indicates that the sum includes only connected diagrams \cite{Bruus}. 
In eq. \ref{Matsubara}, only terms containing products with an even number of phonon operators can give a non-zero contribution.
The lowest non-zero term in $\hat{H}^{\mathrm{e-ph}}$ yields the e-ph self-energy $\Sigma_{n\mathbf{k}}^{\mathrm{e-ph}}$ shown in Fig. 1a, also known as the $GD$ self-energy (in analogy with the $GW$ self-energy \cite{Hybertsen}):
\begin{equation}
\begin{split}
\Sigma_{n \mathbf{k}}^{\mathrm{e-ph}} (i\omega_i,T) = -\frac{1}{\hbar \beta} \frac{1}{\mathcal{N}} \sum_{n' \nu \mathbf{q}} \left| g_{n n' \nu} (\mathbf{k},\!\mathbf{q}) \right|^2 \! \\
\times \sum_j G_{n'\mathbf{k}+\mathbf{q}} (i \omega_i + i \omega_j) \, D_{\nu \mathbf{q}} (i \omega_j)
\label{GD}
\end{split}
\end{equation}
where $\omega_i$ are Matsubara frequencies, and $G$ and $D$ are electron and phonon unperturbed propagators, respectively. 
%
%
Summing over the Matsubara frequencies $\omega_j$ and performing the Wick rotation to the real frequency axis \cite{Bruus}, the imaginary part of the $GD$ diagram gives the e-ph scattering rate in eq. \ref{eph}.\\
%
%
\indent
While elegant, the Matsubara technique can be challenging to apply in practice. A trick to obtain eq. \ref{eph} without summing over Matsubara frequencies is to use non-equilibrium Green's functions on the Keldysh contour combined with the Langreth rules \cite{Jauho}. 
In this technique, the $GD$ self-energy diagram is written as an integral over the Keldysh contour $\mathcal{C}$ \cite{Jauho}: 
\begin{equation}
\Sigma_{n \mathbf{k}}^{\mathrm{e-ph}}  (\omega,T) = \!\! \sum_{n,\nu,\mathbf{q}} \! \left| g_{n n' \nu} (\mathbf{k},\!\mathbf{q}) \right|^2 \!\! \int_{\mathcal{C}} \! d\omega' G_{n'\mathbf{k}+\mathbf{q}} (\omega + \omega') \, D_{\nu \mathbf{q}} (\omega')
\label{Langreth-01}
\end{equation}
and then transformed into a real-axis frequency integral using the Langreth rules \cite{Jauho}:
\begin{equation}
\begin{split}
\!\!\! \Sigma_{n \mathbf{k}}^{\mathrm{e-ph}} (\omega,T) & =  \sum_{n,\nu,\mathbf{q}} \left| g_{n n' \nu} (\mathbf{k},\!\mathbf{q})  \right|^2 \!\! \int_{-\infty}^\infty \! d\omega' \left[ G^{<}_{n'\mathbf{k}+\mathbf{q}} (\omega + \omega') \, D^{A}_{\nu \mathbf{q}} (\omega') \, \right. \\
& \!\!\!\!\!\!\!\!\!\!\!\!\!\!\!\!\!\! \left. + \, G^{R}_{n'\mathbf{k}+\mathbf{q}} (\omega + \omega') \, D^{<}_{\nu \mathbf{q}} (\omega') \right]
\label{Langreth-02}
\end{split}
\end{equation}
After computing the on-shell self-energy by substituting $\omega = E_{n \mathbf{k}}$ and carrying out the integral, the $GD$ e-ph scattering rate in eq. \ref{eph} is obtained through eq. \ref{ex}. The advantage of this contour approach is that the retarded, advanced and lesser Green's functions ($G^R$, $G^A$, and $G^<$ for electrons, respectively, and $D^R$, $D^A$, and $D^<$ for phonons) are tabulated \cite{Jauho}, and the frequency integral is trivial as it merely involves evaluating delta functions. 
This approach is thus a simplified way of deriving eq. \ref{eph}.\\ 
\indent
%
%
One last approach to compute the $GD$ e-ph scattering rate consists in considering the FGR rate of all the processes in which an electron emits or absorbs one phonon, 
and combining them into a first-order rate equation for the time-dependent electron populations $f_{n\mathbf{k}}(t)$: 
\begin{equation}
\begin{split}
\!\!\! \left( \frac{\partial f_{n\mathbf{k}}}{\partial t } \right)^{\mathrm{e-ph}}& = - \frac{2\pi} {\hbar} \frac{1}{\mathcal{N}} \sum_{n' \nu \mathbf{q}} \left| g_{n n' \nu} (\mathbf{k},\!\mathbf{q}) \right|^2 \\ 
& \!\!\!\!\!\!\!\!\!\!\!\!\!\!\!\!\!\!\!\!\!\!\!\!\!\!\!\!\!\!\!\!\!\!\!\! \times \left[ \delta( { \epsilon_{n\mathbf{k}}  - \epsilon_{n' \mathbf{k} + \mathbf{q}} - \hbar \omega_{\nu\mathbf{q}} } ) F_{\mathrm{em}}(t) 
 + \delta ( { \epsilon_{n\mathbf{k}}  - \epsilon_{n' \mathbf{k} + \mathbf{q}} + \hbar \omega_{\nu \mathbf{q}} } ) F_{\mathrm{abs}}(t) \right]
\end{split}
\end{equation}
where the phonon emission ($F_{\mathrm{em}}$) and absorption ($F_{\mathrm{abs}}$) terms are constructed by associating factors $f$ and $1-f$ with the initial and final electronic states in a given scattering process, respectively, and factors $N$ and $N+1$ with phonon absorption or emission:
\begin{equation}
\begin{split}
\!\!\!F_{\mathrm{abs}} &=  f_{n\mathbf{k}} (1 - f_{n'\mathbf{k}+\mathbf{q}}) N_{\nu \mathbf{q}} - f_{n'\mathbf{k}+\mathbf{q}}( 1- f_{n\mathbf{k}}) ( N_{\nu \mathbf{q}} +1) \\
\!\!\!F_{\mathrm{em}} &=  f_{n\mathbf{k}} (1 - f_{n'\mathbf{k}+\mathbf{q}}) (N_{\nu \mathbf{q}} + 1) - f_{n'\mathbf{k}+\mathbf{q}}( 1- f_{n\mathbf{k}}) N_{\nu \mathbf{q}}.
\label{fem}
\end{split}
\end{equation}
For example, the first term in the definition of $F_{\mathrm{abs}}$ corresponds to an electron in the Bloch state $\ket{n\mathbf{k}}$ that scatters into $\ket{n'\mathbf{k}+\mathbf{q}}$ by absorbing one phonon, while the second term is its time-reversal conjugate, the rate of which is added with a minus sign to satisfy the principle of detailed balance. 
%
%
Rewriting the carrier populations as the sum of their value at equilibrium $f^{\mathrm{eq}}_{n\mathbf{k}}$ and the deviation from equilibrium $\delta f_{n\mathbf{k}} (t) = f_{n\mathbf{k}} (t) - f^{\mathrm{eq}}_{n\mathbf{k}}$, and grouping terms that are linear in 
$\delta f_{n\mathbf{k}}(t)$, we obtain:
\begin{equation}
\left( \frac{\partial f_{n\mathbf{k}}}{\partial t } \right)^{\mathrm{e-ph}} = -\frac{\delta f_{n\mathbf{k}}(t)}{\tau^{\mathrm{e-ph}}_{n\mathbf{k}}} + \text{other terms} 
\label{linearization}
\end{equation}
It is seen by carrying out the calculation that $ (\tau^{\mathrm{e-ph}}_{n\mathbf{k}})^{-1}$ in the first term in the righthand side of eq. \ref{linearization} is equal to the e-ph scattering rate of eq. \ref{eph}. The \lq\lq other terms\rq\rq~ are the rate for the scattering processes in which the electrons are scattered back into the state $\ket{n\mathbf{k}}$.
%
%
Neglecting these terms, the solution of eq. \ref{linearization} is an electron population decaying exponentially toward the equilibrium population $f^{\mathrm{eq}}_{n\mathbf{k}}$ with a decay rate $\Gamma_{n\mathbf{k}} = (\tau^{\mathrm{e-ph}}_{n\mathbf{k}})^{-1}$: 
\begin{equation}
f_{n\mathbf{k}} (t) = \left[ f_{n\mathbf{k}}(0) - f^{\mathrm{eq}}_{n\mathbf{k}}\right] e^{-t/\tau^{\mathrm{e-ph}}_{n\mathbf{k}}} + f^{\mathrm{eq}}_{n\mathbf{k}}
\end{equation}
This important result offers a physical interpretation of the $GD$ e-ph scattering rate: it is the \textit{exponential decay rate} of a population of electrons prepared in state $\ket{n\mathbf{k}}$ due to scattering events with phonons. 
Terms that make the population decay deviate from an exponential solution are neglected. This interpretation is consistent with the fact that the perturbed Green's function, through Dyson's equation with the $GD$ self-energy, has the form  
$G_{n\mathbf{k}}(E) = ( E - E_{n\mathbf{k}} + \Sigma_{n \mathbf{k}})^{-1}$, which implies an exponential decay in time of the Fourier transform $G_{n\mathbf{k}}(t)$, with a rate proportional to $\mathrm{Im}\Sigma_{n \mathbf{k}}$. 
The rate-equation model presented above is a particular case of the Boltzmann transport equation for a homogeneous system in the absence of applied fields, as discussed below.\\
\indent
So far, most ab initio calculations of carrier dynamics and transport in the presence of the e-ph interaction have focused on computing the $GD$ self-energy to obtain e-ph RTs, carrier dynamics and transport; additional diagrams (e.g. the Debye-Waller diagram) have been employed to correct the bandstructure due to the e-ph interaction.
It is unclear at present whether higher-order diagrams are necessary to treat cases where the $GD$ diagram approximation is inadequate $-$ for example, at high temperature or for materials with strong e-ph interactions $-$ or whether approaches beyond perturbation theory, such as a strong e-ph coupling theory or a vertex correction, are better suited to model these scenarios. 
First-principles e-ph calculations have not yet ventured into these conceptually more challenging cases.
%
%
\subsection{Electron-electron}
The e-e interaction plays a central role in ground and excited state theories of materials. In the Kohn-Sham equation of DFT, the e-e interactions beyond the Hartree term are approximated by the exchange-correlation potential.  
The latter is a functional of the electronic density \cite{Martin}, and as such it typically lacks an exact expression in terms of Feynman diagrams. 
To obtain accurate quasiparticle energies, and in particular the bandstructure $E_{n\mathbf{k}}$, the Kohn-Sham eigenvalues can be corrected, among other options, through the GW method \cite{Hybertsen}, where $G$ is the electron Green's function, 
and $W$ the screened Coulomb interaction.
In its simplest form, the GW method corresponds to replacing the approximate exchange-correlation functional with a non-local, energy-dependent self-energy,  
which is diagrammatically equivalent to the screened Fock exchange, with the dynamical screening obtained within the random-phase approximation (RPA). 
The quasiparticle states obtained with a rather routine GW calculation, for example using the Yambo \cite{Yambo} or BerkeleyGW \cite{BGW} codes, constitute a well-defined starting point 
to compute e-e scattering processes, as discussed next.\\
\indent
%
%
Interactions among charge carriers (here, electron and hole quasiparticles) are mediated by the screened Coulomb interaction, $W$, and cause a finite carrier lifetime, defined here as the e-e RT, $\tau_{n\mathbf{k}}^{\mathrm{e-e}}$. Phonon or impurity assisted e-e processes are also possible \cite{Ridley,Takeshima-02}, in which additional momentum in the scattering process is provided by a phonon or impurity, respectively.
%
%
The two main e-e scattering mechanisms discussed here are the Auger and impact ionization (IIZN) processes \cite{Govoni,Kotani,Ridley} (see Fig. 1b). Auger scattering and IIZN are critical to understanding carrier dynamics in materials, as they determine the recombination \cite{Govoni} and energy loss rate \cite{Bernardi-Si} of excited carriers, especially at high carrier concentrations and for carriers with large energy excess with respect to the band edges. In particular, Auger scattering limits the efficiency of electronic, optoelectronic and photovoltaic devices through carrier recombination.\\
\indent
%
%
In Auger processes, an electron-hole pair recombines and gives out the energy to a third carrier, which is an electron in so-called $eeh$ processes and a hole in $hhe$ processes. The Auger mechanism can thus equivalently be seen as the scattering of an electron pair ($eeh$ process) or a hole pair ($hhe$ process) into final states, with the restriction that an electron-hole pair recombines, thus changing the number of carriers.\footnote{Intraband Auger and IIZN processes that do not change the number of carriers are also possible, though not discussed here. Some of these processes can be computed using out-of-equilibrium GW \cite{Marini-KBE}.} %
%
While complex analytical treatments of Auger scattering employing two-particle, three-particle \cite{Cini} and even four-particle\cite{Takeshima} Green's functions have been explored (especially for processes involving core states), first-principles Auger calculations have so far focused on valence electron processes and employed relatively simple theory, as discussed below.\\
\indent
%
%
%
The IIZN is the inverse process of Auger recombination (see Fig. 1b). In IIZN scattering, a carrier loses energy transitioning to a state energetically closer to the band edges, within the conduction or valence bands for electrons and holes respectively. Energy is conserved by exciting an electron-hole pair, across the band gap in a semiconductor and across the Fermi energy in a metal. At finite temperature, Auger and IIZN rates need to be computed together and combined, since they are related by detailed balance.\\
\indent
First-principles calculations of Auger and IIZN scattering can build on the quasiparticle bandstructure and dynamical screened Coulomb interaction computed within GW. However, to reduce computational cost most first-principles Auger calculations carried out so far employed either model dielectric screening \cite{Picozzi,Steiauf,Kioupakis-03} or the static (zero frequency) RPA dielectric screening \cite{Govoni}. On the other hand, IIZN calculations employing the dynamical 
screened Coulomb interaction are rather common.\\
\indent
At zero temperature, the imaginary part of the GW self-energy, $ \mathrm{Im}\Sigma^{\mathrm{GW}}$, computed on-shell at the quasiparticle energy, can be employed to compute the IIZN scattering rate:\\
\begin{equation}
\Gamma_{n\mathbf{k}}^{\mathrm{e-e, IIZN}} = (\tau_{n\mathbf{k}}^{-1})^{\mathrm{e-e, IIZN}} = \frac{2}{\hbar} \mathrm{Im}\Sigma^{\mathrm{GW}}(\omega=E_{n\mathbf{k}})
\label{GW}
\end{equation}
This rate accounts for the creation of electron-hole pairs upon energy loss by a single quasiparticle. Such IIZN calculations using GW have been employed extensively to interpret ARPES linewidths and electronic RTs \cite{Kotani,Rubio-metals}.
For example, we recently employed GW calculations to compute the IIZN contribution to the RTs of excited carriers in semiconductors and metals \cite{Bernardi-Si,Bernardi-GaAs,Bernardi-Au}.\\
\indent
%
%
Auger processes, both direct (namely, mediated by the screened Coulomb interaction alone) and phonon assisted, have been computed from first principles chiefly in simple semiconductors \cite{Govoni,Steiauf,Kioupakis,Kioupakis-03}.
%
%
The key quantities in first-principles calculations of Auger scattering are the matrix elements of the screened Coulomb interaction. 
Using the shorthand notation $j_1 \equiv (n_1 \mathbf{k}_1)$ to label the band and crystal momentum of Bloch states, Auger scattering can be seen as the scattering from initial states $j_1$ and $j_2$ into final states $j_3$ and $j_4$ 
(see Fig. 1b). 
The screened Coulomb interaction $W_{1234}$ among the four Bloch states involved in Auger scattering has the well-known expression:
%
%
\begin{equation}
\begin{split}
W_{1234} & \equiv \braket{\psi_{j_1} \psi_{j_2} | W | \psi_{j_3} \psi_{j_4}} \\
& \!\!\! = \frac{1}{V} \sum_{\mathbf{G}\mathbf{G'}} \rho_{n_1,n_3}(\mathbf{k_1}, \mathbf{q}, \mathbf{G}) W_{\mathbf{G} \mathbf{G'}}(\mathbf{q}) \rho^{*}_{n_4,n_2}(\mathbf{k_4}, \mathbf{q}, \mathbf{G'}) \\
& \!\!\! \times \delta_{\mathbf{k_3},\mathbf{k}-\mathbf{q}+\mathbf{G}}\, \delta_{\mathbf{k_4},\mathbf{k}+\mathbf{q}+\mathbf{G'}}
\label{w1234}
\end{split}
\end{equation}
where $\mathbf{q}$ is the transferred momentum folded to the first BZ, and 
$\rho_{n,n'}(\mathbf{k}, \mathbf{q}, \mathbf{G}) = \braket{n\mathbf{k} | e^{i(\mathbf{q} + \mathbf{G}) \mathbf{r}}  | n'\mathbf{k}-\mathbf{q}}$ are dipole matrix elements.
%
%
The screened Coulomb interaction in a plane-wave basis, $W_{\mathbf{G} \mathbf{G'}}(\mathbf{q})$, depends on the approximation employed for the screening. 
For the RPA screening used in GW calculations, it reads:
\begin{equation}
W_{\mathbf{G} \mathbf{G'}}(\mathbf{q}) = v(\mathbf{q}+\mathbf{G})\delta_{\mathbf{G},\mathbf{G'}} + v(\mathbf{q}+\mathbf{G}) \chi^0_{\mathbf{G} \mathbf{G'}}(\mathbf{q},\omega) v(\mathbf{q}+\mathbf{G'})
\label{RPA}
\end{equation}
where $v(\mathbf{q} + \mathbf{G})=4\pi e^2 / \left| \mathbf{q} + \mathbf{G} \right|^2$ is the bare Coulomb interaction, and $\chi^0_{\mathbf{G} \mathbf{G'}}(\mathbf{q},\omega)$ the frequency-dependent RPA polarization function \cite{Hybertsen,Yambo}.
%
%
Auger scattering rate computations require a large number of screened Coulomb interaction matrix elements. To reduce computational cost, the static (i.e., zero frequency) polarization $\chi^0_{\mathbf{G} \mathbf{G'}}(\mathbf{q},\omega=0)$ has been employed in a few works \cite{Govoni}, though a common approach is to approximate the static dielectric screening with a model function, $\epsilon^M(\mathbf{q},\omega=0)$, in which case:
\begin{equation}
\!\!\!\!\!\! W^{M}_{\mathbf{G} \mathbf{G'}}(\mathbf{q}) = \frac{1}{\epsilon^M(\mathbf{q}+\mathbf{G})} \frac{4\pi e^2}{ \left| \mathbf{q}+\mathbf{G} \right|^2 + \lambda^2} \delta_{\mathbf{G},\mathbf{G'}}
\end{equation}
where $\lambda$ is the Debye screening length (e.g., due to free carriers). While employed widely in GW calculations, the dynamical RPA screening in eq. \ref{RPA} has not been employed to study Auger scattering to our knowledge.\\
\indent
%
%
The Auger rate (AR) can be computed using the FGR, by obtaining the square e-e matrix elements $\left| M_{1234} \right|^2$ for Auger scattering among four given electronic states. Care must be taken to take spin into account correctly. 
For processes in which states $j_1$ and $j_2$ possess the same spin, both the direct term $W_{1234}$ and the exchange term $W_{1243}$ need to be included:
%
%
\begin{equation}
\left| M^{\uparrow \uparrow}_{1234} \right|^2 = \left| W_{1234} - W_{1243} \right|^2
\label{same-spin}
\end{equation}
%
%
while for the case of states $j_1$ and $j_2$ with opposite spin, the square matrix elements of the direct and exchange processes are summed \cite{Ridley}:
\begin{equation}
\left| M^{\uparrow \downarrow}_{1234} \right|^2 = \left| W_{1234}\right|^2 + \left| W_{1243} \right|^2
\label{same-spin}
\end{equation}
The total matrix element involved in Auger scattering for any given four states is thus:
\begin{equation}
\left| M_{\mathrm{1234}} \right|^2 = \left| M^{\uparrow \uparrow}_{1234} \right|^2 + \left| M^{\uparrow \downarrow}_{1234} \right|^2
\label{eeme}
\end{equation}
Finally, the AR can be expressed using the FGR in terms of the \textit{total} rate $R^{\mathrm{AR}}$ for any carrier to recombine through Auger processes. 
This AR corresponds to the rate of change of the carrier population per unit volume, and includes a factor of 2 to sum over spin:
\footnote{To avoid double-counting, the sum in eq. \ref{AR} must be carried out over distinct pairs of initial and final states, as opposed to over all states $j_i$.} 
\begin{equation}
\!\!\!\! R^{\mathrm{AR}} = 2 \frac{2\pi}{\hbar} \sum_{\mathrm{j_1\, j_2\, j_3\, j_4}} \mathcal{P} \left| M_{\mathrm{1234}} \right|^2 \delta(E_{j_1} + E_{j_2}  - E_{j_3} - E_{j_4})
\label{AR}
\end{equation}
where the population factor is $\mathcal{P} = f_1 f_2 (1-f_3) (1-f_4)$ for $eeh$ processes and $\mathcal{P} = (1-f_1) (1- f_2) f_3 f_4$ for $hhe$ processes.
Alternatively, the \textit{quasiparticle} AR of a carrier in a given Bloch state with a given spin can also be computed:
\begin{equation}
\!\!\!\! \Gamma^{\mathrm{e-e, AR}}_{j_1} = \frac{2\pi}{\hbar} \sum_{\mathrm{j_2\, j_3\, j_4}} \mathcal{P}' \left| M_{\mathrm{1234}} \right|^2 \delta(E_{j_1} + E_{j_2} - E_{j_3} - E_{j_4})
\label{AR-qp}
\end{equation}
where $\mathcal{P}' = \mathcal{P} / f_1 $ for $eeh$ processes and $\mathcal{P}' = \mathcal{P} / (1-f_1) $ for $hhe$ processes.\\
\indent
To compute the ARs, fine BZ grids are needed to converge the sums over initial and final momenta \cite{Govoni}, similar to the case of e-ph scattering. 
Wannier interpolation of the band structure has been recently employed to achieve such fine BZ sampling \cite{Kioupakis,Steiauf,Kioupakis-03}. 
Recent work \cite{Kioupakis-03} has extended the formalism discussed above to compute phonon-assisted ARs, by combining model dielectric screening with ab initio e-ph matrix elements.
%
%
So far, first-principles AR calculations have been carried out only for simple semiconductors, often employing model screening rather than the dynamical RPA screening. 
Possible future directions include employing dynamical screening, computing phonon or defect assisted processes, and treating more complex materials.
%
%
\subsection{Phonon-phonon}
Phonon dispersions and eigenvectors can be routinely computed from first principles within the harmonic approximation, using either DFT plus finite-differences or density functional perturbation theory (DFPT) \cite{Baroni}. 
While phonons in the harmonic approximation possess an infinite lifetime, scattering with electrons and other phonons results in finite phonon RTs.
The latter can be measured with Raman or neutron scattering experiments, or computed from first principles using DFPT \cite{Deinzer}, among other approaches (e.g, finite differences, molecular dynamics, etc).
The phonon lifetime due to scattering with phonons, defined here as the ph-ph RT, $\tau^{\mathrm{ph-ph}}_{\nu \mathbf{q}}$, can be obtained from the imaginary part of the ph-ph self-energy, $\mathrm{Im}\Sigma^{\mathrm{ph-ph}}_{\nu \mathbf{q}}$, through eq. \ref{phx}. The lowest-order ph-ph self-energy is shown in Fig. 1c.\\
\indent
To obtain the ph-ph RTs, the total energy of the crystal $E(\{\mathbf{u}_{is}\})$ is first expanded as a Taylor series in the small ion displacements $\mathbf{u}_{is}$ about the equilibrium positions \cite{Cowley-01,Cowley-02}, with $s$ the ion and $i$ the lattice site labels: 
%
%
%
\begin{equation}
\begin{split}
&E(\{\mathbf{u}_{is}\}) \!\!=\!\! E_0 + \frac{1}{2} \sum_{\substack{ i_1 i_2 \\ s_1 s_2 \\ \alpha_1 \alpha_2}} \phi^{(2)}(i_1 s_1 \alpha_1; i_2 s_2 \alpha_2) \mathbf{u}_{i_1 s_1 \alpha_1} \mathbf{u}_{i_2 s_2 \alpha_2} \\
& + \!\frac{1}{3!} \!\!\!\! \sum_{\substack{ i_1 i_2 i_3\\ s_1 s_2 s_3\\ \alpha_1 \alpha_2 \alpha_3}} \!\!\! \phi^{(3)}(i_1 s_1 \alpha_1; i_2 s_2 \alpha_2; i_3 s_3 \alpha_3) \mathbf{u}_{i_1 s_1 \alpha_1} \mathbf{u}_{i_2 s_2 \alpha_2} \mathbf{u}_{i_3 s_3 \alpha_3} \\
& + \!\mathcal{O}(\{\mathbf{u}_{is}\}^4)
\end{split}
\label{en}
\end{equation}
%
where $E_0$ is the ground-state energy, the second term containing the $\phi^{(2)}$ coefficients gives the harmonic approximation and the remainder are the anharmonic terms. 
The $n$-th order coefficients $\phi^{(n)}$ are equal to the $n$-th derivatives of the total energy with respect to the ion displacements about the equilibrium positions. 
For example, the coefficients $\phi^{(2)}(i_1 s_1 \alpha_1; i_2 s_2 \alpha_2)$ are the force constants which give the force in the $\alpha_1$ direction on the atom $s_1$ in the unit cell $i_1$ when atom $s_2$ in the unit cell $i_2$ is displaced in the $\alpha_2$ direction. 
The harmonic approximation then consists of neglecting the terms containing more than two displacements and diagonalizing the resulting Hamiltonian to give the phonon dispersions and eigenvectors.\\
\indent
The anharmonicity of the crystal potential about the lattice equilibrium positions, given by the terms in eq. \ref{en} containing three or more displacements, causes phonons to scatter with each other \cite{Cowley-01}. 
In third-order processes, which are regulated by the force constants $\phi^{(3)}$, two phonons can merge to form one with higher frequency (upconversion process), or a phonon can split into two lower-frequency phonons (difference process). 
The third-order ph-ph RTs computed from first principles can capture such three-phonon decay processes \cite{Debernardi-01,Debernardi-02,Vanderbilt}. 
Higher orders, such as four-phonon processes, are more challenging to compute, and are chiefly relevant at high temperature, in phase transitions, or for thermal expansion studies \cite{Fultz}. Such quartic terms have rarely been computed 
using first-principles calculations.\\
\indent
The anharmonic ph-ph scattering rates, $\Gamma_{\nu \mathbf{q}}^{\mathrm{ph-ph}} = (\tau^{\mathrm{ph-ph}}_{\nu \mathbf{q}})^{-1}$, are obtained from eq. \ref{en} by substituting the second-quantized form of the lattice displacements $\mathbf{u}_{is}$ (see eq. \ref{displacement}). 
Third-order anharmonic coupling matrix elements $V^{(3)}(\nu_1 \mathbf{q_1}; \nu_2 \mathbf{q_2}; \nu_3 \mathbf{q_3})$ are then defined, which quantify the coupling between three modes with branch indices $\nu_i$ and crystal momenta $\mathbf{q}_i$, with $i=1,2,3$. The expression for $V^{(3)}$ involves generalized lattice Fourier transforms of the force constants $\phi^{(3)}$, and is given in ref. \cite{Cowley-01}. 
A finite-temperature perturbation expansion of the phonon propagator in powers of $V^{(3)}$ is thus developed, leading to ph-ph processes to all orders of perturbation theory \cite{Cowley-01,Cowley-02}. 
The final result for the third order ph-ph scattering rate for a phonon with a given branch index $\nu$ and wavevector $\mathbf{q}$ is:
%
%
\begin{equation}
\begin{split}
\Gamma_{\nu \mathbf{q}}^{\mathrm{ph-ph}} & = \frac{18 \pi}{ \hbar^2} \sum_{\substack{\nu_1 \nu_2 \\ \mathbf{q}_1 \mathbf{q}_2}} \left| V_{\mathrm{ph}}^{(3)}( \nu \mathbf{q}, \nu_1 \mathbf{q}_1, \nu_2 \mathbf{q}_2) \right|^2  \!\! \delta_{\mathbf{q} + \mathbf{q}_1 + \mathbf{q}_2,\mathbf{G}}  \\
& \times \left[ (N_{\nu_1 \mathbf{q}_1} \!+\! N_{\nu_2 \mathbf{q}_2} \!+\! 1) \delta(\omega_{\nu \mathbf{q}} - \omega_{\nu_1 \mathbf{q}_1} - \omega_{\nu_2 \mathbf{q}_2}) \right. \\
& \left. + 2 (N_{\nu_1 \mathbf{q}_1} \! - \! N_{\nu_2 \mathbf{q}_2}) \delta( \omega_{\nu \mathbf{q}} + \omega_{\nu_1 \mathbf{q}_1} - \omega_{\nu_2 \mathbf{q}_2} ) \right]
\end{split}
\label{phph}
\end{equation}
where $N_{\nu \mathbf{q}}$ are Bose-Einstein phonon populations, $\omega_{\nu \mathbf{q}}$ are phonon frequencies, the factor of $18$ arises from the ways of pairing up the phonon lines in the self-energy diagram \cite{Cowley-02}, and the Kronecker delta expresses crystal momentum conservation. The first term in the square bracket describes upconversion processes, and the second difference processes.\\ 
\indent
Alternative ways of deriving eq. \ref{phph} include the rate equation and contour integral techniques discussed above. 
The rate-equation derivation is particularly simple, and worth discussing briefly. In this approach, the rate of change of the phonon population is written as a sum of the FGR rates of three-phonon processes. 
Each participating phonon is either absorbed, with an associated population factor of $N$, or emitted, with an associated population factor of $N+1$. Writing the FGR for all the three-phonon processes, 
associating to them the coupling matrix elements $V_{\mathrm{ph}}^{(3)}( \nu \mathbf{q}, \nu_1 \mathbf{q}_1, \nu_2 \mathbf{q}_2)$, and using a multiplicity factor $\mathcal{M}=9$ that accounts for process combinatorics, we obtain:
\begin{equation}
\begin{split}
&\!\!\!\!\! \left(\frac{\partial N_{\nu \mathbf{q}}} { \partial t} \right)^{\!\!\mathrm{ph-ph}} = - \frac{2\pi}{\hbar^2}  \mathcal{M} \sum_{\substack{\nu_1 \nu_2 \\ \mathbf{q}_1 \mathbf{q}_2}} \left| V_{\mathrm{ph}}^{(3)}( \nu \mathbf{q}, \nu_1 \mathbf{q}_1, \nu_2 \mathbf{q}_2) \right|^2 \times \\
&\!\!\!\!\!\! \times\!\! \{\left\{ \left[ N_{\nu \mathbf{q}} (N_{\nu_1 \mathbf{q}_1} \!+\! 1) ( N_{\nu_2 \mathbf{q}_2} \!+\! 1) \right]  \!-\! \left[ N_{\nu_1 \mathbf{q}_1} N_{\nu_2 \mathbf{q}_2} (N_{\nu \mathbf{q}}\!+\!1) \right] \right\} \!\delta(\omega \!-\! \omega_1 \!-\! \omega_2 ) \\ 
&\!\!\!\!\!\! +\! \left\{ \left[ N_{\nu \mathbf{q}} N_{\nu_1 \mathbf{q}_1} ( N_{\nu_2 \mathbf{q}_2} \!+\! 1) \right]  \!-\! \left[ N_{\nu_2 \mathbf{q}_2} (N_{\nu \mathbf{q}}\!+\!1) (N_{\nu_2 \mathbf{q}_2}\!+\!1) \right] \right\} \!\delta(\omega \!+\! \omega_1 \!-\! \omega_2 ) \\
&\!\!\!\!\!\!\! +\! \left\{ \left[ N_{\nu \mathbf{q}} N_{\nu_2 \mathbf{q}_2} ( N_{\nu_1 \mathbf{q}_1} \!+\! 1) \right]  \!-\! \left[ N_{\nu_1 \mathbf{q}_1} (N_{\nu \mathbf{q}}\!+\!1) (N_{\nu_1 \mathbf{q}_1}\!+\!1) \right] \right\} \!\delta(\omega \!-\! \omega_1 \!+\! \omega_2 ) \}
\label{crazy}
\end{split}
\end{equation}
where each distinct three-phonon process is enclosed in a square bracket, and carries a plus or minus sign according to whether it involves absorption or emission of the $\ket{\nu \mathbf{q}}$ phonon, respectively.\footnote{The $\hbar^2$ in the denominator derives from using frequency instead of energy conservation.} We simplify eq. \ref{crazy} by rewriting the phonon populations as $N_{\nu \mathbf{q}} = N^{\mathrm{eq}}_{\nu\mathbf{q}} + \delta N_{\nu \mathbf{q}}$, namely, as the sum of the equilibrium populations and the deviation from equilibrium, and separate the terms linear in $\delta N_{\nu \mathbf{q}}$:
\begin{equation}
\left(\frac{\partial N_{\nu \mathbf{q}}} { \partial t} \right)^{\!\!\mathrm{ph-ph}} = - \frac{\delta N_{\nu \mathbf{q}}(t)}{ \tau^{\mathrm{ph-ph}}_{\nu \mathbf{q}} } + \text{other terms}
\label{phph-bte}
\end{equation}
It is seen by inspection that $(\tau^{\mathrm{ph-ph}}_{\nu \mathbf{q}})^{-1}$ is equal to the ph-ph scattering rate of eq. \ref{phph}. We thus interpret the ph-ph RT 
as the exponential decay lifetime of a population of phonons with a given branch index and wavevector, due to three-phonon scattering processes.
This treatment neglects deviations from an exponential decay as well as fourth and higher order diagrams related to scattering processes involving four or more phonons, respectively.\\
\indent
%
%
Ph-ph scattering processes play a key role in phonon dynamics. For example, the ph-ph interaction is the main contribution to the RT of a phonon in intrinsic semiconductors and insulators. 
Accordingly, ph-ph scattering regulates the thermal conductivity, and it further contributes in all materials (including metals) to determine electron dynamics, and in particular the equilibration rate of excited carriers \cite{Shah}. 
First-principles calculations of ph-ph RTs have enabled dramatic advances in microscopic understanding of phonon dynamics, as outlined below.\\
\indent
%
%
\subsection{Phonon-electron}
As a result of the interaction with electrons, phonons acquire a finite lifetime, the so-called ph-e RT, $\tau_{\nu \mathbf{q}}^{\mathrm{ph-e}}$, which plays a central role in phonon transport in metals and doped semiconductors. 
It should not be confused with the e-ph RT discussed above.
The ph-e scattering rate, $\Gamma_{\nu \mathbf{q}}^{\mathrm{ph-e}}$, can be computed through eq. \ref{phx} from the imaginary part of the lowest-order ph-e self-energy (see Fig. 1d):  
\begin{equation}
\begin{split}
\Gamma_{\nu \mathbf{q}}^{\mathrm{ph-e}} (T) &= \frac{2}{\hbar}  \mathrm{Im}\Sigma^{\mathrm{ph-e}}_{\nu \mathbf{q}} \! = 2 \frac{2\pi}{\hbar} \frac{1}{\mathcal{N}} \sum_{n n' \mathbf{k}}\! \left| g_{n n' \nu}(\mathbf{k},\mathbf{q}) \right|^2 \\
&\,\, \times \left( f_{n\bf{k}} - f_{n'\mathbf{k}+\bf{q} } \right) \delta( \epsilon_{n\mathbf{k}}  \! - \epsilon_{n'\mathbf{k} \! + \mathbf{q}} + \hbar\omega_{\nu \mathbf{q}} )
\end{split}
\label{phe}
\end{equation}
In the lowest-order of perturbation theory, ph-e scattering processes arise as phonons polarize the electron gas by generating electron-hole pairs with a center-of-mass momentum $\mathbf{q}$, in a process mediated by the e-ph interaction. 
Accordingly, the ph-e self-energy diagram consists roughly of the RPA polarization function $\chi^0$ multiplied by the square of the e-ph coupling matrix elements, i.e., $\Sigma^{\mathrm{ph-e}}_{\nu \mathbf{q}} \approx g^2 (\mathbf{q})\chi^0(\mathbf{q}, \omega_{\nu \mathbf{q}})$.
The possible approaches to derive the ph-e scattering rate in eq. \ref{phe} within perturbation theory are analogous to those discussed above for the e-ph case. For example, using the rate-equation approach, the rate of change of the phonon population  $N_{\nu \mathbf{q}}$ due to scattering with electrons reads: 
\begin{equation}
\begin{split}
\left( \frac{N_{\nu \mathbf{q}}}  {\partial t}  \right)^{\mathrm{\!\!ph-e}} =& -\frac{4 \pi}{\hbar} \frac{1}{\mathcal{N}} \sum_{n n'\bf{k}} \left| g_{n n' \nu}(\mathbf{k},\mathbf{q}) \right|^2 \\
&  \times \delta({\epsilon_{n\mathbf{k}}  - \epsilon_{n' \mathbf{k}+\mathbf{q}} + \hbar \omega_{\nu \mathbf{q}} }) F_{\mathrm{abs}}
\end{split}
\label{bte-phe} 
\end{equation}
where $F_{\mathrm{abs}}$ has been defined above in eq. \ref{fem}. 
%
%
%
Analogous to the ph-ph case, we can write:
\begin{equation}
\left( \frac{N_{\nu \mathbf{q}}} {\partial t}  \right) ^{\mathrm{\!\!ph-e}} \!= -\frac{ \delta N_{\nu \mathbf{q}}(t)}{ \tau_{\nu \mathbf{q}}^{\mathrm{ph-e}} } + \text{other terms}
\label{phe-02}
\end{equation}
%
%
where it is seen by carrying out the calculation that $(\tau_{\nu\mathbf{q}}^{\mathrm{ph-e}})^{-1}$ in eq. \ref{phe-02} is equal to the ph-e scattering rate in eq. \ref{phe}. 
The ph-e RT thus defines the exponential decay constant for a phonon population initially occupying a given phonon mode with defined branch index and wavevector, due to the interaction with electrons within the lowest order of perturbation theory. 
Similar to the e-ph case, terms that make the population decay deviate from an exponential are neglected in eq. \ref{phe}. Among other applications, \mbox{ph-e} RTs have been combined with ph-ph RTs in advanced studies of phonon dynamics \cite{Bonini,Bonini-02}.
\section{First-principles carrier dynamics}
\label{sec:2}
The framework presented above centers on computing scattering rates and RTs of carriers and phonons. Using these quantities, dynamics and transport in materials can be studied from first principles. 
The Boltzmann transport equation (BTE) is the theoretical approach underlying much of the first-principles work in carrier and phonon dynamics. 
In this section, we first discuss the BTE, and then briefly mention recent developments on first-principles implementations of the Kadanoff-Baym equations (KBEs), an approach that can include quantum dynamical effects 
not captured by the BTE.
%
We close the section by discussing applications to dynamics and transport. 
The key contributions of first-principles calculations in this area include understanding dynamical and excited state processes with unprecedented microscopic detail $-$ thus both complementing experiment and providing deeper understanding $-$ as well as guiding the development of novel technologies in electronics, lighting, and energy conversion and storage.
%
We highlight trends in recent research and discuss several example calculations, without attempting to provide a comprehensive review.
%
%
%
\subsection{Boltzmann transport equation}
The BTE describes the flow in phase space of the electron and phonon occupations, $f_{n \mathbf{k}}(\mathbf{r},t)$ and $N_{\nu \mathbf{q}}(\mathbf{r},t)$, respectively. 
These occupations can be interpreted as the probability distributions of an electron (phonon) occupying a state with given crystal momentum and band (branch) index, at coordinate $\mathbf{r}$ and time $t$. Since the crystal momentum and position are both specified at the same time, the occupation distributions, and thus the BTE, are semiclassical in nature. 
The first-principles formalism extends standard analytical treatments \cite{AM} by accounting for multiple electronic bands and phonon branches, and employing materials properties computed without empirical parameters.\\
\indent
The dynamics of electrons and phonons in the BTE is typically split into two parts, a slowly varying flow in coordinate and momentum space that is driven by external fields and commonly called \lq\lq drift\rq\rq, 
and a collision dynamics induced by scattering processes, which leads to discrete transitions in electron and phonon momentum space \cite{Smith}. The time evolution of the electron and phonon occupations are thus the sum of a drift 
and a collision flow. 
%
%
The BTE is adequate to describe length and time scales spanning multiple scattering events, i.e., the so-called diffusive or hydrodynamic regime. It further assumes the validity of Fermi liquid theory, and thus that 
interacting quasiparticles such as carriers and phonons possess a one-to-one mapping to their non-interacting counterparts. Accordingly, interactions are seen as collisions among the quasiparticles, the scattering rates of which are 
computed using first-principles many-body perturbation theory.\\
\indent
The BTE for electrons in the presence of a force field $\mathbf{F}$ reads \cite{Mahan,Smith}:
\begin{equation}
\!\!\!\! \frac{\partial f_{n\mathbf{k}}(\mathbf{r},t) }{\partial t } \!=\! - \left[ \mathbf{\nabla}_{\mathbf{r}}f_{n\mathbf{k}}(\mathbf{r},t) \!\cdot\! \mathbf{v}_{n\mathbf{k}} \!+\! \hbar^{-1} \mathbf{\nabla}_{\mathbf{k}} f_{n\mathbf{k}}(\mathbf{r},t) \!\cdot\! \mathbf{F} \right] + \mathcal{I}[f_{n\mathbf{k}}]
\label{BTE}
\end{equation}
where $\mathbf{v}_{n\mathbf{k}}$ are band velocities, the first bracket gives the drift term, and the collision term is given by the scattering integral $\mathcal{I}[f_{n\mathbf{k}}]$, which is a functional of the electron populations:
\begin{equation}
\begin{split}
\mathcal{I} =& - \sum_{n'\mathbf{k'}} \Gamma_{n\mathbf{k} \rightarrow n' \mathbf{k'}} \cdot f_{n\mathbf{k}} (\mathbf{r},t) \left[ 1 - f_{n'\mathbf{k'}} (\mathbf{r},t) \right] \\
&+ \sum_{n\mathbf{k}} \Gamma_{n'\mathbf{k'} \rightarrow n \mathbf{k}} \cdot f_{n'\mathbf{k'}} (\mathbf{r},t) \left[ 1 - f_{n\mathbf{k}} (\mathbf{r},t) \right] 
\end{split}
\label{scattering}
\end{equation}
Here, $\Gamma_{n\mathbf{k} \rightarrow n' \mathbf{k'}}$ are scattering rates from the electronic Bloch state $\ket{n\mathbf{k}}$ to other states $\ket{n'\mathbf{k'}}$, and the first line gives the total scattering rate out of the state $\ket{n\mathbf{k}}$, while the second line gives the scattering from all other states to $\ket{n\mathbf{k}}$. Typical driving forces in the BTE include electric and magnetic fields, strain, and temperature or chemical potential gradients \cite{Mahan,Smith}.
The material properties enter the BTE both in the drift term through the bandstructure and in the collision term through the scattering rates.\\ 
\indent
To simplify the scattering integral, the relaxation time approximation (RTA) can be introduced, in which the electron population returns to equilibrium with a rate proportional to the deviation 
$\delta f_{n\mathbf{k}} (t) = f_{n\mathbf{k}}(t) - f^{\mathrm{eq}}_{n\mathbf{k}}$ from the equilibrium population $f^{\mathrm{eq}}_{n\mathbf{k}}$. Within the RTA, the scattering integral in eq. \ref{scattering} becomes: 
\begin{equation}
\begin{split}
\mathcal{I}[f_{n\mathbf{k}}] = - \frac{\delta f_{n\mathbf{k}} (t)}{\tau_{n\mathbf{k}}}
\label{RT}
\end{split}
\end{equation}
In the absence of applied forces and for a homogeneous system, so that the drift term vanishes, the so-called state-dependent RTA thus yields: 
\begin{equation}
\frac{\partial f_{n\mathbf{k}}(t)} {\partial t} = - \frac{\delta f_{n\mathbf{k}}(t)} {\tau_{n\mathbf{k}}}
\end{equation}
i.e., the equation found above for the case of e-ph scattering.
The solution of the state-dependent RTA is a carrier population in which each state decays exponentially in time to the equilibrium population:
\begin{equation}
f_{n\mathbf{k}}(t) = \left[ f_{n\mathbf{k}}(0) - f^{\mathrm{eq}}_{n\mathbf{k}} \right] e^{-t/\tau_{n\mathbf{k}}} + f^{\mathrm{eq}}_{n\mathbf{k}}
\end{equation}
\indent 
Most first-principles calculations for carriers out of equilibrium rely on some form of RTA, and have so far been limited to the case of homogeneous systems in which the spatial dependence of the carrier population is neglected. 
One limitation of the RTA approach is that it does not conserve the number of particles and energy, so that it is adequate only to obtain rough timescale estimates in specific conditions (e.g., low carrier density).
A current focus of first-principles calculations is the computation of carrier transport coefficients, which involve solving the BTE at steady-state under an applied field. 
Transport properties such as electrical and thermal conductivities and thermoelectric coefficients can be computed either within the RTA or with numerical solutions of the BTE that explicitly include the scattering integral. 
The BTE can be solved beyond the RTA using iterative \cite{Fugallo,WuLi} and Monte Carlo approaches, among others.\\
\indent
A BTE for phonons can also be derived:
\begin{equation}
\!\!\! \frac{\partial N_{\nu\mathbf{q}}(\mathbf{r},t) }{\partial t } \!=\! -\! \left[ \mathbf{\nabla}_{\mathbf{r}}N_{\nu\mathbf{q}}(\mathbf{r},t) \!\cdot\! \mathbf{v}_{\nu\mathbf{q}} + \hbar^{-1} \mathbf{\nabla}_{\mathbf{q}} N_{\nu\mathbf{q}}(\mathbf{r},t) \cdot \mathbf{F} \right]  
+ \mathcal{I}[N_{\nu\mathbf{q}}]
\label{BTE-ph}
\end{equation}   
where $\mathbf{F}$ is a driving force (e.g., strain or a temperature gradient), the phonon velocities $\mathbf{v}_{\nu\mathbf{q}}$ are obtained from the phonon dispersions, and $\mathcal{I}[N_{\nu\mathbf{q}}]$ is the scattering integral. The scattering integral including three-phonon processes and isotopic scattering is given in refs. \cite{Smith,Fugallo}.
Considerations analogous to the electron BTE also hold for phonons, including the wide use of the RTA to compute transport coefficients such as the phonon thermal conductivity. Studies of thermal transport that employ first-principles data as input to solve the phonon BTE in inhomogeneous (e.g., nanoscale) systems are also being actively investigated \cite{Mingo,Romano}.\\
\indent
%
When considering e-ph coupling, the electron and phonon scattering rates depend on both the electron and phonon occupations, thus coupling the electron and phonon BTEs. 
%
%
In the absence of driving fields, the solution for interacting electrons and phonons would involve a large system of coupled differential equations, with size $\mathcal{N}_{\mathrm{e}} = \mathcal{N}_b \times \mathcal{N}_k$ for electrons and $\mathcal{N}_{\mathrm{ph}} = \mathcal{N}_{\nu} \times \mathcal{N}_{\mathbf{q}}$ for phonons, where $\mathcal{N}_b$ and $\mathcal{N}_{\mathbf{k}}$ are the number of bands and $\mathbf{k}$-points, respectively, and $\mathcal{N}_{\nu}$ and $\mathcal{N}_{\mathbf{q}}$ the number of phonon branches and $\mathbf{q}$-points. 
Solving such a large set of coupled differential equations would be challenging, but it may enable dramatic advances 
in understanding coupled carrier and phonon dynamics, a holy grail in solid state physics. Coupled ab initio electron-phonon dynamics could be applied to a variety of important scenarios in device physics and spectroscopy. 
Given the recent progress in computing electron and phonon dynamics individually, it appears that the time is ripe for tackling coupled electron-phonon dynamics from first principles.\\
%
%
%
\subsection{Kadanoff-Baym equations}
As highlighted in the previous section, the Boltzmann transport theory is semiclassical, and further assumes the validity of perturbation and Fermi liquid theory.
The BTE is a useful tool to investigate a variety of materials and transport properties, and enables wide-ranging studies of electron and phonon dynamics.
%
However, the BTE describes scattering processes in an incoherent regime, and as such it cannot be employed to study the dynamics of coherent superpositions of states. 
In addition, an ansatz needs to be made about the initial nonequilibrium electron and phonon populations. For the important case of ultrafast dynamics following a short laser pulse excitation, 
the change in carrier (or phonon) populations due to both the incipient excitation and scattering cannot be included on the same footing in the BTE, though in reality the two processes often overlap in time.\\
\indent
To overcome these and other limitations, quantum kinetic equations have been developed that can more rigorously describe electron and phonon dynamics out of equilibrium \cite{Kadanoff}.
The Kadanoff-Baym equations (KBEs) are an example of quantum kinetic equations based on non-equilibrium Green's functions. 
We will discuss the KBEs only very briefly, chiefly because they represent an emerging area of investigation in first-principles dynamics; for more comprehensive discussions, see refs. \cite{Mac-ne,Stefanucci,Kadanoff,Stan}.\\
\indent
The KBEs have been developed and traditionally applied in the context of model systems described by simplified Hamiltonians \cite{Jauho}. 
Recent numerical implementations of the KBEs for inhomogeneous systems \cite{Stan} have brought the approach one step closer to the first principles community.
In these numerical implementations, the external fields are treated non-perturbatively and the many-body interactions are included through a $\Phi$-derivable self-energy that guarantees the 
macroscopic conservation laws of the system are fulfilled \cite{Kad}.\\
\indent 
Next, we provide some basic facts about the KBEs. Using the compact notation $1 = (\mathbf{r}_1,t_1)$ for space-time coordinates, the contour-ordered Green's function satisfies the equation of motion:
\begin{equation}
i\partial_{t_1} G(1,2) = h(1) G(1,2) + \delta(1,2) + \int_{\mathcal{C}} \!d3\, \Sigma(1,3)G(3,2)
\label{KBE}
\end{equation}
where $h(1)$ is the local part of the Hamiltonian, the self-energy $\Sigma$ captures all the interactions beyond the Hartree, and $\mathcal{C}$ denotes integration on the Keldysh contour. 
Some of the approximations employed so far for $\Sigma$ include the Hartree-Fock (HF), the GW, and the second Born approximation.  
Expanding the equation of motion in component Green's functions for different parts of the contour, and employing the Langreth rules \cite{Jauho} to convert integrals containing the product of functions on the contour to integrals on the real time axis, eq. \ref{KBE} is converted to multiple dynamical equations, called the KBEs. The KBEs describe the flow of the non-equilibrium Green's functions defined on different tracks of the Keldysh contour, including the retarded, advanced, lesser, greater, mixed vertical, and Matsubara Green's functions \cite{Stan}. 
For example, the KBE for the lesser Green's function reads:
\begin{equation}
i\partial_{t} G^{<}(t,t') = h^{\mathrm{HF}}(t) G^{<}(t,t') + \mathcal{I}_{\mathrm{KBE}}^{<}[\Sigma](t,t')
\label{lesser}
\end{equation}
where we suppressed the coordinates and introduced the HF Hamiltonian $h^{\mathrm{HF}}$ and the scattering integral $\mathcal{I}_{\mathrm{KBE}}^{<}[\Sigma]$, defined as a functional of the self-energy (for details, see ref. \cite{Stan}). 
A full set of KBEs can de derived for the other relevant non-equilibrium Green's functions \cite{Stan}. 
%
%
For a specific self-energy approximation, the KBEs together with the initial conditions fully determine the non-equilibrium Green's functions at all times.\\
\indent 
Solution of the KBEs yields, through the non-equilibrium Green's functions, dynamical quantities of interest such as the electron and phonon time-dependent populations.
%
%
In practice, numerical solution of the KBEs is computationally expensive because it requires time-stepping the non-equilibrium Green's functions with the \textit{two times} $t$ and $t'$ on the real axis.
The presence of two times stems from the fact that memory effects are included in the KBEs and that the equations include coherent effects, but it makes the method computationally expensive and limits the longest timescales that can be accessed. 
A first principles implementation, for example with a plane-wave basis set, would make challenges related to computational cost even more severe. On the other hand, because the two-time KBEs can include coherent effects, they have the potential to describe physics beyond the Fermi liquid theory approximation.
%
%
Recently, Sangalli et al. \cite{Marini-KBE,Sangalli} developed and applied a first-principles version of the KBEs that employs the so-called completed collision approximation to reduce the non-equilibrium Green's function 
dynamics to one time variable. 
%
The formalism employs the dynamical RPA dielectric screening together with ab initio e-ph and e-e scattering rates. 
Their method, together with other first-principles implementations of the KBEs currently under investigation by several groups, have the potential to become accurate tools to study electron and phonon dynamics in materials. 
%
%
\subsection{Electron and phonon transport}
First-principles calculations of transport properties are being actively investigated in the electronic structure community, as they constitute at present an important missing link between ab initio microscopic and device-scale models. 
For example, the electrical conductivity $\sigma$ and charge mobility $\mu$ can be computed from first principles within the BTE formalism. The linearized BTE at steady state gives \cite{Mahan,BoltzWann}: 
\begin{equation}
\sigma_{\alpha\beta} = e^2 \sum_{n\mathbf{k}} \tau_{n\bf{k}} (\mathbf{v}_{n\mathbf{k}})_\alpha (\mathbf{v}_{n\mathbf{k}})_\beta \left(-\partial f / \partial E \right)  
\label{sigma}
\end{equation}
where $\mathbf{v}_{n\mathbf{k},\alpha}$ is the $\alpha$-th component of the band velocity for the Bloch state $\ket{n\mathbf{k}}$, obtained from DFT or GW bandstructures, $\tau_{n\mathbf{k}}$ are carrier RTs, and $f(T)$ is the temperature dependent Fermi-Dirac distribution. The mobility $\mu$ can be obtained from the conductivity using $\mu = \sigma/(ec)$, where c is the carrier concentration and $\sigma$ the direction-averaged conductivity. By varying the chemical potential, both the electron and hole mobilities can be computed.\\
\indent
In state-of-the-art calculations, eq. \ref{sigma} is evaluated by interpolating the bandstructure to obtain the band velocities on fine BZ grids, and computing the RTs for scattering with phonons or through other processes. 
This approach yields phonon-limited transport properties, which are relevant in relatively pure crystalline materials at room temperature.
Scattering with defects, either elastic \cite{Lordi} or inelastic \cite{Wang-nonrad}, is important in many cases of practical relevance and has also been computed from first principles, though work in this area is still in its nascent stage.\\
\indent
The BoltzWann \cite{BoltzWann} and BoltzTrap \cite{BoltzTrap} codes implement conductivity and mobility calculations, and can interpolate the bandstructure with maximally localized Wannier functions \cite{W90-code} and Fourier interpolation, respectively. The state-dependent RTs are often approximated as a constant or a slowly varying function of energy. The constant RT can be used as a fitting parameter or extracted from experiment, thus making the calculation semi-empirical. First-principles calculations of the RTs we recently developed \cite{Bernardi-Si,Bernardi-GaAs,Bernardi-Au,Bernardi-transport} reveal a non-trivial dependence of the RTs on band and $\mathbf{k}$-point, which is not captured by the constant RT approximation and improves the agreement of the computed conductivity with experiment \cite{Bernardi-transport}. 
For example, we recently employed first-principles calculations with state-dependent RTs to compute the room temperature resistivity of the three noble metals Cu, Ag, and Au, 
and obtained agreement within 5\% of experiment.
%
%
This work further showed that the e-ph relaxation times vary significantly on the Fermi surface (e.g., by up to a factor of 3 in Cu), contrary to the conventional wisdom that constant RTs are a good approximation in metals.
The RTs were found to correlate with the Fermi surface topology and the orbital character of the electronic states. Due to computational cost, transport calculations employing ab initio RTs have so far been carried out mainly on materials with simple unit cells; research efforts to compute RTs in larger system are underway.\\ 
\indent
%
%
For organic materials and correlated oxides, first-principles calculations of carrier mobility are still very challenging. The narrow width of the electronic bands induces strong e-ph coupling, and the carriers in these materials, also known as polarons, become localized and strongly coupled with the lattice. Boltzmann transport theory (e.g., eq. \ref{sigma}) cannot be applied to compute the conductivity due to the localized nature of the carriers, and the e-ph interaction often cannot be treated perturbatively. Transport in organic materials and correlated oxides is typically described as a hopping process of localized polarons, leading to a peculiar temperature dependence of the mobility. Transport calculations in organic semiconductors and correlated oxides from first principles constitute a rapidly growing field.\\
\indent
%
%
The carrier diffusion length, namely the distance carriers travel before recombining, is one of the most important parameters in applications, e.g., to design photovoltaic and photoelectrochemical active layers. The diffusion length could be obtained from first principles by calculating both the carrier mobility and the recombination lifetimes due to radiative \cite{Bernardi-exciton} and non-radiative \cite{Wang-nonrad} processes.
However, such diffusion length calculations are still nearly absent in the literature, mainly due to the challenge of including multiple recombination processes. 
Computing and validating carrier recombination lifetimes remains an open challenge for first-principles calculations.\\
\indent
%
%
Auger scattering is the recombination mechanism most widely studied from first principles, especially in simple elemental and III$-$V semiconductors\cite{Govoni,Kioupakis,Steiauf,Kioupakis-03}. For example, Auger processes in GaN and related nitride compounds employed in lighting applications have been studied recently \cite{Kioupakis,Steiauf,Kioupakis-03}, along with Auger rates in ionic crystals employed in scintillators \cite{Scintillators}. 
While there has been exciting progress in this area, Auger processes are still complex to compute quantitatively since they can be mediated by the Coulomb interaction alone, or they can be phonon or defect assisted. The interactions to include in these calculations are challenging to establish a priori, and since measuring Auger rates experimentally is difficult, validation of the computed results is still non-trivial.\\
\indent
Heat transport, and in particular the thermal conductivity of solids due to both electrons and phonons, can be computed from first principles using the ph-ph and ph-el RTs given above. We touch on this topic only very briefly. 
Solving the phonon BTE at steady state within the RTA provides the expression for the lattice thermal conductivity $\kappa$ employed in most first-principles calculations:
\begin{equation}
\kappa_{\alpha \beta} = \frac{\hbar^2}{\mathcal{N} V_{uc} k_B T^2} \sum_{\nu \mathbf{q}} (\mathbf{v}_{\nu \mathbf{q}})_{\alpha} (\mathbf{v}_{\nu \mathbf{q}})_{\beta} \, \omega^2_{\nu \mathbf{q}} \, N_{\nu \mathbf{q}} (N_{\nu \mathbf{q}} + 1) \tau_{\nu \mathbf{q}}
\label{thermal}
\end{equation}
where the phonon velocities $\mathbf{v}_{\nu \mathbf{q}}$ and the RTs can be computed from first principles. For cases in which both the ph-ph and ph-e interactions are important, the respective RTs are combined using Matthiessen's rule \cite{Mahan}, i.e., the scattering rates for the different processes are added up to give the total scattering rate.\\
\indent
%
%
In thermal transport, the ph-ph scattering processes are typically divided into normal (N) and Umklapp (U); isotopic phonon scattering can also be included \cite{broido-01,broido-02,Cepellotti}. For N processes, all three phonon momenta belong to the first BZ, whereas in U processes the sum of the three momenta equals a reciprocal lattice vector. Such distinction is key for heat transport, since N processes do not dissipate heat as a consequence of energy and momentum conservation, while U processes act as the source of intrinsic dissipation.
%
%
While employing the state-dependent phonon RTs in eq. \ref{thermal} provides good agreement with experiment for some materials \cite{broido-01}, iterative solution of the phonon BTE (see refs. \cite{Fugallo,Mingo}) typically improves agreement with experiment and is necessary for quantitative accuracy in some cases \cite{Cepellotti}. 
Similar iterative solutions are also being explored for the electron BTE \cite{WuLi}. First-principles thermal transport computations are providing unprecedented insight into phonon RTs \cite{broido-01,broido-02,Cepellotti} and ballistic mean free paths \cite{Minnich,Minnich-02}. New design rules for materials with high or low thermal transport are being investigated, with the potential to greatly advance applications in electronics and thermoelectrics \cite{Minnich-03}.
%
\subsection{Excited carrier dynamics}
The framework presented above enables novel studies of excited carrier dynamics, a topic of particular relevance given the recent experimental advances in ultrafast lasers and spectroscopy. 
This section focuses on recent work by the author on ultrafast dynamics and hot carriers (HCs), namely, excited carriers with excess energy with respect to the band edges. 
HCs are an important source of energy loss in solar cells \cite{Bernardi-Si} and light-emitting devices \cite{Kioupakis}. Collecting HCs before they equilibrate could greatly enhance the efficiency of solar cells.
For example, for the case of Si under standard solar illumination, nearly 25\% of incident solar energy is lost to heat as the HCs generated by sunlight absorption thermalize to the edges of the band gap. 
Experimentally, this thermalization process is difficult to control and understand with microscopic detail due to the sub-ps time scale involved.\\
\indent 
%
%
We recently studied HCs in solar cells by combining first-principles calculations of e-ph and e-e scattering with the BTE \cite{Bernardi-Si}. 
We demonstrated that a HC distribution characteristic of Si under solar illumination thermalizes within 350 fs, in excellent agreement with pump-probe experiments \cite{Bernardi-Si}. 
%
%
The work further employed first principles calculations to obtain the average distances traveled by carriers before losing energy through phonon emission $-$ the so-called ballistic mean free paths (MFPs) $-$ in semiconductors \cite{Bernardi-Si} and metals \cite{Bernardi-Au}. 
The MFPs define the limit thickness to extract HCs from a device before they lose energy, and are important quantities given that the idea of extracting HCs to increase the energy conversion efficiency is widely pursued in photovoltaic and photoelectrochemical devices. The MFPs along different crystallographic directions can be computed by multiplying the band velocity by the RT, namely, $ L_{n\mathbf{k}} =  (\mathbf{v}_{n\mathbf{k}}\cdot \hat{\mathbf{k}} ) \, \tau_{n\mathbf{k}}$, where the unit crystal momentum vector $\mathbf{\hat{k}}$ is oriented along the ballistic propagation direction. 
%
%
Our computed MFPs of 10$-$15 nm for electrons and holes in Si \cite{Bernardi-Si} are in excellent agreement with recent scanning tunneling microscopy experiments \cite{Lock}. The MFPs computed for noble metals such as Au and Ag show a volcano shape with maximum MFPs of 15$-$30 nm, which rapidly degrade to $\sim$5 nm for HCs with energies a few eV away from the Fermi energy. These trends suggest that extracting HCs with a few eV excess energy may require 
very thin metallic layers, and at metal-semiconductor interfaces HC extraction may be challenging due to the disordered interface layer typical of metal-semiconductor junctions with thickness comparable with the MFPs.\\
\indent
%
%
The e-ph RTs of HCs in GaAs were also investigated, focusing on excited electrons occupying the $\Gamma$, $L$, and $X$ conduction valleys  \cite{Bernardi-GaAs}. 
The first-principles e-ph scattering rates in GaAs were found to be in excellent agreement with values obtained by fitting experimental transport data \cite{Bernardi-GaAs}. 
This work also contributed to resolve a controversy in the interpretation of ultrafast optical experiments in GaAs, demonstrating unambiguously that the tens of femtoseconds decay times observed experimentally arise from e-ph scattering. 
By computing the scattering rate due to different phonon modes, our work showed that, contrary to common notions, all optical and acoustic modes contribute substantially to excited electron energy loss, with a dominant contribution from transverse acoustic phonons.\\
\indent
%
%
The conventional wisdom that excited carriers equilibrate chiefly by emitting longitudinal optical (LO) phonons needs to be revisited. 
This traditional picture is based on the notion that the Fr{\"o}hlich interaction, which couples electrons to the polar LO mode with an e-ph matrix element $g(q) \!\propto\! 1/q$ ($q$ is the absolute value of the phonon wavevector), dominates in materials with polar bonds such as GaAs.
However, the Fr{\"o}hlich interaction is mainly active near the band edges, where intraband scattering with small transferred momentum $q$ prevails. 
At higher energy where interband transitions dominate the phase space, we found in GaAs and several other polar materials that acoustic phonon emission is the dominant mechanism for HC energy loss, and that in general multiple phonon modes contribute substantially to energy dissipation.
The ability to resolve the contribution to carrier dynamics of different electronic bands and phonon modes is truly unique to first-principles calculations, which should be seen as an advanced computational spectroscopy tool that can greatly extend the scope of ultrafast experiments.
A more comprehensive review of first principles HC dynamics will be presented elsewhere.\\
\indent
%
%
Finally, recent advances in first-principles dynamics using the KBE \cite{Marini-KBE,Sangalli} enable studies that are not possible with the BTE. For example, the possibility to include the light pulse in the dynamical equations while coherently 
propagating the carriers is crucial to model experiments in which the pulse duration is of the same order of magnitude (e.g., $\sim$100 fs) as the dynamics of interest. 
The KBE approach further enables studies of carrier dynamics under relatively high laser intensities, for which the RTA breaks down due to lack of energy and carrier number conservation.
We note that even advanced treatments such as the KBE currently rely on perturbation theory with fixed nuclei positions. 
However, treating the nuclei as fixed is a severe approximation for a material irradiated by intense laser light. 
Extensions will be necessary to properly include nuclear dynamics in approaches based on many-body perturbation theory such as the BTE and KBE. 
Real-time time-dependent DFT \cite{Octopus}, on the other hand, allows for nuclear motion through a variety of schemes, and remains at present the tool of choice to model materials irradiated by intense lasers. 
Given the recent trend of employing lasers with growing intensities (e.g., the free-electron laser \cite{FEL}) for materials spectroscopy, the boundary between different spectroscopy communities is blurring. New opportunities are arising as a result for first-principles calculations to contribute to the future of ultrafast spectroscopy.
\vspace{-10pt}
\section{Conclusion}
First principles calculations of carrier and phonon dynamics are emerging as rigorous extensions of the ground-state DFT and excited-state GW-BSE methods. 
They can tackle fundamental problems in solid-state physics and help develop novel technological applications, thus redefining the boundaries of ab initio theories. 
Looking forward, the unique ability of first-principles calculations to microscopically interpret spectroscopy and transport experiments will play a key role to bridge the gap between increasingly complex 
experiments and their microscopic interpretation.
In particular, understanding electron, phonon and spin transport with microscopic detail will fuel the development of novel electronics in the post Moore's law era, solid-state renewable energy devices, and novel spectroscopy techniques. 
The next decade of first-principles calculations will be at the heart of these exciting developments.
\vspace{-15pt}
\section{Acknowledgement}
The author thanks Jamal Mustafa, Luis Agapito, and Jin-Jian Zhou for fruitful discussions, and Davide Sangalli, Vatsal Jhalani, Bolin Liao and Celene Barrera for feedback on the manuscript. 
This work was supported by a start-up fund from the California Institute of Technology. 

\end{document}